# Characterizing Hydration of SDS Micelles by Contrast Variation Small Angle Neutron Scattering

*Katherine Chen,[†] Chi-Huan Tung,[†‡#] and Changwoo Do[\*,†]*

[†]Neutron Scattering Division, Oak Ridge National Laboratory, Oak Ridge, Tennessee 37831, United States

[‡]Department of Materials Science and Engineering, National Tsing Hua University, Hsinchu 30013, Taiwan

[#]Shull Wollan Center, The University of Tennessee and Oak Ridge National Laboratory, Oak Ridge, Tennessee 37831, USA

**ABSTRACT:** Small-angle neutron scattering (SANS) from cationic globular micellar solutions composed of sodium dodecyl sulfate (SDS) and in water was studied with contrast variation approach. Extensive computational studies have demonstrated that the distribution of invasive water is clearly an important feature for understanding the self-organization of SDS molecules and the stability of assemblies. However, in existing scattering studies the degree of hydration level was not examined explicitly. Here using the scheme of contrast variation, we establish a methodology of SANS to determine the intra-micellar radial distributions of invasive water and SDS molecules from the evolving spectral lineshapes caused by the varying isotopic ratio of water. A detailed description hydration of SDS micelles is provided, which in an excellent agreement with known results of many existing simulations studies. Extension of our method can be used to provide an in-depth insight into the micellization phenomenon which is commonly found in many soft matter systems.

Sodium dodecyl sulfate (SDS) are amphiphilic molecules consisting of a sulfate group as the polar head and a hydrophobic dodecyl group in the tail region. In the aqueous environment they are known to self-organize into a variety of aggregates with different geometric shapes depending on the external thermodynamic conditions. SDS micelles are among the most thoroughly studied ionic micellar systems and has been treated as a model micellar system to provide insight into the links between the molecular structure, interaction, composition and self-assembled conformation and properties.

Extensive computational studies have examined the structure and dynamics of SDS micelles in great detail.[1–18] Areas of research include micellar size, shape, aggregation number, and their dependence on counterion association and temperature,[1,2,12] radial distributions and dynamics of invasive water, counterions and SDS molecules,[4,13,18] thermal stability and kinetics and diffusivity of surfactants exchange, and transport properties of associated surfactants.[9,10] On the other hand, much experimental activity, mainly small angle neutron scattering (SANS), has also been devoted to the structure study of SDS micelles to determine geometry and size of an individual micelle and inter-micellar interactions.[19–28]

On the context of computation simulation[1–18] and theoretical investigation[29–36], the conformational description usually begins with the consideration of water penetration. The radial distribution of invasive water population is calculated under different ionic conditions and temperatures to elucidate the role of hydrophobic effect, and its interplay with the electrostatic repulsion among the headgroups region, in determining the self-assembly processes of SDS molecules. Based on the structure of SDS molecule, SDS micelles have been treated as core-shell particles with impervious surface in existing SANS studies.[19–28] This modeling restricts the examination of water profile from the analysis of scattering data. Experimental studies of invasive water distribution within a SDS micelle are scarce so far. This challenge provides the motivation of our study.

The present work advances the current understanding about micellization of SDS molecules in the following ways: using SANS technique complemented by the scheme of contrast variation, we are able to identify the intra-micellar water and SDS molecular distributions, two fundamental spatial correlation functions which determine the conformation of SDS micelles. These two distribution functions allow us to calculate the conformational parameters of SDS micelles and to understand hydration structure of SDS micelles.

Sodium dodecyl sulfate (SDS) surfactants were commercially obtained from Sigma and used without further purification. Deuterium oxide ($D_2O$) was purchased from Cambridge Isotope Laboratories, Inc. Samples with a fixed SDS concentration of 50 mg/mL were prepared. The aqueous solvents used in this experiment were prepared by mixing a predetermined amount of $D_2O$ and de-ionized water with molar ratios of $D_2O$ to $H_2O$, defined as $\gamma$, at 100:0, 90:10, 80:20, 70:30, 60:40, 50:50 and 40:60. Samples were prepared by weighing the appropriate mass of surfactant and salt and adding the necessary volume of water with different values of $\gamma$ to achieve the desired concentrations. SANS measurements were conducted at the Extended Q-Range Small-Angle Neutron Scattering Diffractometer (EQ-SANS) at the Spallation Neutron Source (SNS) at Oak Ridge National Laboratory (ORNL). A sample-to-detector distance of 1.3m with an incident neutron wavelength of 6 Å was used to cover a $Q$-range of 0.01 to 0.3 Å$^{-1}$. The SDS micellar samples were accommodated by banjo cells with a path length of 1 mm, and all the measurements were performed at 25 °C using a thermal bath temperature control module. Measured scattering intensities were corrected for detector background, sensitivity, and empty cell scattering, and were normalized to absolute units using the porous silica standard sample as described elsewhere.[37,38]

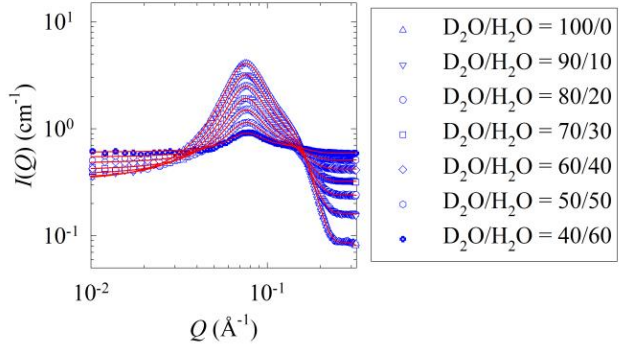

**Figure 1.** The SANS intensity distribution $I(Q)$ (symbols) and the associated model fitting curves (solid curves) of SDS micellar solutions for seven different scattering contrasts. Quantitative agreement between the experimental data and model curves is seen. The experimental uncertainties are on the order or smaller than the symbol size.

Figure 1 shows the SANS absolute intensity distributions $I(Q)$ and the corresponding model fitting curves obtained from SDS micellar solutions at seven different levels of scattering contrast (D$_2$O/H$_2$O = 100/0, 90/10, 80/20, 70/30, 60/40, 50/50 and 40/60). The peak of $I(Q)$ centering around 0.08 Å$^{-1}$ is an indication of the strong electrostatic interactions between neighboring micelles. A steady increase in the molar ratio of H$_2$O in the solvent deteriorates the scattering contrast between SDS micelles and background water and results in a progress decrease in the scattering intensity.

Our goal is to extract the intra-micellar water and SDS surfactant distribution using SANS complemented by the technique of contrast variation. The measured SANS absolute intensity $I(Q)$ is generally modelled by the following factorization expression:

$$I(Q) = n_m \Delta\rho_m^2 v_m^2 P(Q)S(Q) + I_{inc}, (1)$$

where $n_m$ is the micellar number density, $\Delta\rho_m$ is the difference of the scattering length density (SLD) between particles and background solvent, $v_m$ is the volume of a SDS micelle, and $P(Q)$ is the form factor which describes the intra-micellar spatial correlation, $S(Q)$ is the inter-micellar structure factor which conceals the information regarding the spatial distribution of micelles in solutions, and $I_{inc}$ the incoherent background, which shows no dependence on scattering wavevector $Q$. Instrument resolution is also considered during the data analysis process.

In this work, Yukawa potential is used to model the inter-micellar screened Coulomb interaction $V(r)$. Given this specific mathematical expression of $V(r)$, $S(Q)$ is obtained via solving the Ornstein-Zernike (OZ) integral equation (OZ) with the modified penetrating-background corrected rescaled mean spherical approximation (MPB-RMSA) closure.[36] Evidenced by the existing scattering studies,[24,27] conformation of SDS micelles can be phenomenologically modelled by a fuzzy concentric core-shell sphere and the corresponding $P_m(Q)$ can be expressed as

$$P_m(Q) = |F_m(Q)|^2$$
$$= [r_c F_1(Q, R_1, \sigma_1) + (1 - r_c)F_2(Q, R_1, \sigma_g)]^2, (2)$$

where $F_1(Q, R, \sigma) = 3j_1(QR)\exp(-\sigma^2 Q^2/4)/(QR)$ and $j_1(QR)$ is the spherical Bessel function of the first kind of order 1. $R_1$ is the radius of hydrophobic core consisting of dodecyl groups. $\sigma_1$ is a parameter measuring the fuzziness of the interface between hydrophobic core and hydrophilic corona. Computation simulations of SDS micelles in water have shown that the sulfate groups are essentially localized around the micellar periphery. Moreover, its scattering length density (SDS) is much higher than that of the dodecyl tail. We therefore use a gamma distribution $F_2(Q, a, b) = \frac{\sin[(a-1)\phi]}{(1+b^2Q^2)^{\frac{a-1}{2}}(a-1)bQ}$ where $a = \left(\frac{R_1}{\sigma_g}\right)^2$, $b = \frac{\sigma_g^2}{R_1}$ and $\phi = \tan^{-1}(bQ)$ to describe the radial distribution of the sulfate groups. $\sigma_g$ is used to quantify the spatial fluctuation of sulfate groups caused by the invasive water and association of lithium anions. $r_c$ and $1 - r_c$ are the weighting factors quantifying the relative scattering contribution of each term.

Before presenting the details of SANS data analysis, a brief description of our approach is helpful to depict the strategy for extracting the conformational heterogeneity of SDS micelles from SANS experiment: Driven by the entropic effect, computational studies have demonstrated that water molecules surrounding a SDS micelle will penetrate into the micellar interior especially the hydrophilic corona region. Due to the confinement effect, the packing pattern and SLD of invasive water are different from those of bulk water. As a result, both SDS surfactants and invasive water contribute to the measured coherent scattering intensity via compositional and density differences in comparison to the bulk water background. Therefore, both should be considered as the constituent components of an individual SDS micelle and the measured $I(Q)$ is a collective reflection of the spatial arrangements of SDS surfactants and invasive water. If their spatial distributions do not depend on $\gamma$, by altering the D/H ratio of solvent, the scattering contributions from SDS molecules and invasive water can be uniquely compartmentalized from the characteristic variation of $I(Q)$ presented in Figure 1 with the additional boundary conditions provided by the variation of $\gamma$.

Eqn. (2) clearly shows that the evolution of $I(Q)$ in Figure 1 is caused by the change in $\Delta\rho_m$. This observation offers a convenient starting point for formulating the framework of data analysis. An explicit expression of $\Delta\rho_m$ and $\gamma$ is required to address the intra-micellar mass heterogeneity. Given the spherical symmetry, one can express $\Delta\rho_m$ as

$$\Delta\rho_m = \int \Delta\rho_m(r) 4\pi r^2 dr, (3)$$

where $\Delta\rho_m(r)$ is the intra-micellar radial SLD distribution. The range of integration is the volume of a micelle. From Eqns. (2) and (3), one can express the coherent scattering power of a micelle as

$$\Delta\rho_m^2 v_m^2 P_m(Q) = \iint dr dr' \Delta\rho_m(r) \Delta\rho_m(r') \exp[-i\mathbf{Q} \cdot (\mathbf{r} - \mathbf{r}')], (4)$$

The range of integration is same as that in Eqn. (3). From Eqns. (2) and (4), one can show that $\Delta\rho_m(r)$ takes the following expression

$$\Delta\rho_m(r) = \Delta\rho_m v_m [r_c f_1(r, R_1, \sigma_1) + (1 - r_c) f_2(r, R_1, \sigma_g)], (5)$$

where $f_1(r, R_1, \sigma_1) = \frac{3}{8\pi R^3}\left\{ erf\left(\frac{r+R_1}{\sigma_1}\right) - erf\left(\frac{r-R_1}{\sigma_1}\right) + \frac{\exp\left[-\frac{(r+R_1)^2}{\sigma_1^2}\right] - \exp\left[-\frac{(r-R_1)^2}{\sigma_1^2}\right]}{\sqrt{\pi} r} \right\}$ and $erf(x)$ is the error function of $x$. $f_2(r, R, \sigma) = \frac{r^{a-1}}{4\pi\Gamma(a+2)b^{a+2}} e^{-\frac{r}{b}}$ and $\Gamma(x)$ is the gamma function of x. By analyzing the SANS spectra given in Figure 1 using Eqns. (1) and (2), all the parameters in Eqn. (5) can be uniquely determined except for $\Delta\rho_m v_m$. However, because $P_m(0) = 1$, from Eqn. (1) the numerical values of $n_m \Delta\rho_m^2 v_m^2$ at different $\gamma$ can be determined using the amplitude of $I(Q)$. By multiplying both sides of Eqn. (5) by $(n_m)^{1/2}$, the quantity $\Delta\rho_m(r)(n_m)^{1/2}$ can be are experimentally obtained from Eqns. (1) and (2).

$\Delta\rho_m(r)$ given in Eqn. (3) can be further expressed as the difference between the SLD distribution of a micelle $\rho_m(r)$ and that of the background water $\rho_w = b_w/v_w$, where $b_w$ is the scattering length of bulk water and $v_w$ is its molecular volume (30 Å$^3$). One can define $\rho_m(r) \equiv \rho_p(r) + \rho_w(r)$ where $\rho_p(r)$ and $\rho_w(r)$ are the radial SLD distributions of SDS molecules and invasive water respectively. Accordingly, $\Delta\rho_m(r)$ can be alternatively expressed as

$$\Delta\rho_m(r) = \rho_p(r) + \rho_w(r) - \rho_w$$
$$= \rho_p(r) + b_w \left[H(r) - \frac{1}{v_w}\right], (6)$$

where $H(r)$ is the radial number density distributions of invasive water. Given $b_w = \gamma b_{D_2O} + (1-\gamma) b_{H_2O}$, $\Delta\rho_m(r)(n_m)^{1/2}$ can be expressed as a linear function of $\gamma$. Namely

$$\Delta\rho_m(r,\gamma)(n_m)^{1/2}$$
$$= \gamma(b_{D_2O} - b_{H_2O})(n_m)^{1/2}\left[H(r) - \frac{1}{v_w}\right]$$
$$+ (n_m)^{1/2}\left\{\rho_p(r) + b_{H_2O}\left[H(r) - \frac{1}{v_w}\right]\right\}, (7)$$

Eqns. (3) and (7) provides the explicit connection between $\Delta\rho_m$ and $\gamma$. Now we can quantitatively address the intra-micellar mass heterogeneity from the dependence of $\Delta\rho_m(r)(n_m)^{1/2}$ on $\gamma$. By differentiating Eqn. (7) with respect to $\gamma$, only the first term on the RHS of Eqn. (7) remains. After rearranging the coefficients, it is found that

$$(n_m)^{1/2}\left[H(r) - \frac{1}{v_w}\right] = \frac{1}{b_{D_2O} - b_{H_2O}} \frac{d}{d\gamma}[\Delta\rho_m(r,\gamma)(n_m)^{1/2}], (8)$$

From the experimentally determined $\Delta\rho_m(r,\gamma)(n_m)^{1/2}$, the LHS of Eqn. (8) can be determined with linear regression on $\gamma$. It is important to note that Eqn. (8) only determines the value of the product of $(n_m)^{1/2}$ and $H(r) - \frac{1}{v_w}$, and thus one more boundary condition is required to uniquely determine both $n_m$ and $H(r)$. Due to the strong hydrophobicity of dodecyl group, computational studies have demonstrated that the core region of SDS micelles is not accessible by invasive water. This observation allows us to specify $H(0) = 0$ and accordingly $n_m$ can be determined from Eqn. (8). Moreover, since SDS molecules take up all the intra-micellar interior space which is not accessible by invasive water, the radial number density distributions of SDS molecules $P(r)$ can be determined from the following expression

$$P(r) = -\frac{\left[H(r) - \frac{1}{v_w}\right] v_w}{v_{SDS}}, (9)$$

where $v_{SDS}$ is the volume of a SDS molecule defined in Eqn. (1). Based on its molecular weight ($m_{SDS} = 265$ g/mol) and density ($d_{SDS} = 1.01$ g/cm$^3$), $v_{SDS}$ is found to be 437.3 Å$^3$. Given the radial distributions of water and SDS populations, the volume fraction distribution $f_w(r)$ and $f_{SDS}(r)$ within $r$ to $r + dr$ can be calculated by multiplying $H(r)$ and $P(r)$ by $v_w$ and $v_{SDS}$. Moreover, one can estimate the volume of sulfate head group from $v_{SDS}$ and the volume of dodecyl $v_{tail}$. Since the molecular weight of dodecyl $m_{tail} = 170.34$ g/mol and the density $d_{tail} = 0.749$ g/cm$^3$, $v_{tail}$ is found to be 379 Å$^3$. Therefore, $v_{SO_4} = v_{SDS} - v_{tail} = 58.3$ Å$^3$. Accordingly, one can estimate the volume fraction distribution of sulfate headgroup $f_{head}(r)$ by multiplying the normalized gamma distribution $\Delta\rho_2(r, R, \sigma)$ with the micellar aggregation number given later in Figure 3. The results of $f_w(r)$, $f_{SDS}(r)$ and $f_{head}(r)$ are given in Figure 2.

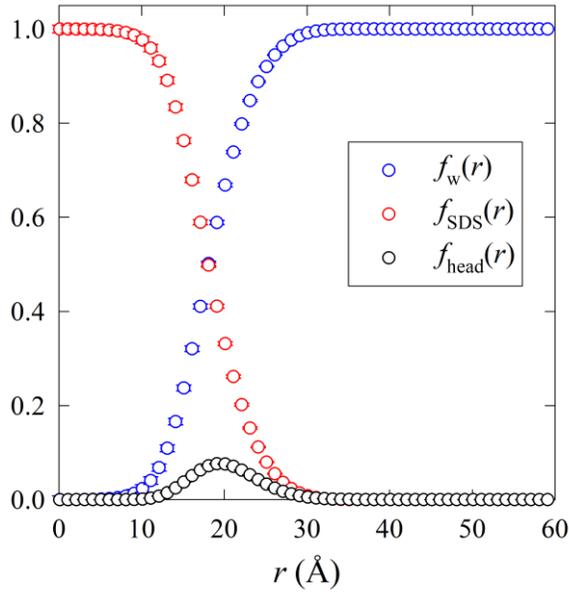

**Figure 2.** The radial distributions of volume fraction of intra-micellar water ($f_w(r)$, blue symbols), SDS molecules ($f_{SDS}(r)$, red symbols), and headgroup ($f_{head}(r)$, black symbols).

Figure 2 provides the radial distributions of invasive water molecules ($f_w(r)$), SDS molecules ($f_{SDS}(r)$) and head groups ($f_{head}(r)$) in terms of volume fraction to quantitatively describe the intra-micellar heterogeneous mass distributions. When $r > 50$ Å, all the $f_w(r)$ are seen to reach the asymptotic value of 1. Now one can use $f_w(r)$ extracted from this experimental method to directly gauge the computationally predicted population profile of invasive water. The micellar central region ($r < 10$Å) occupied by dodecyl tails essentially remains unhydrated. This observation is consistent with the computational and SANS results.[4–6,9,12,18,20–22]

One can the intra-micellar space $v_{ew}$ which is not accessible by water from $H(r)$. Namely,

$$v_{ew} = -v_w \int_0^\infty \left[H(r) - \frac{1}{v_w}\right] 4\pi r^2 dr, (10)$$

Accordingly, the aggregation number $N_{agg}$ can be defined as

$$N_{agg} = \frac{v_{ew}}{v_u} - 1, (11)$$

$N_{agg}$ is found to be approximately 70, which is in good agreement with existing literature value. One can also calculate $N_{agg}$ from $P(r)$ and the results are found to be identical to those calculated from Eqns. (10) and (11). The number density of SDS micelles $n_m$ can be further calculated to evaluate the dependence of micellization on salt concentration. Given the boundary condition of $H(0) = 0$, from Eqn. (8) it is found that

$$n_m = \left\{\frac{v_w}{b_{D_2O} - b_{H_2O}} \frac{d}{d\gamma}\left[\Delta\rho_m(r)(n_m)^{1/2}\right]\right\}^2, (12)$$

The calculated $n_m$ is given as $1.54 \times 10^{-6}$ Å$^{-3}$. Given the weight fraction of SDS molecules in solution (50 mg/mL), the number fraction of SDS molecules associated in micelles, $f_m$, can be calculated from $n_m$ and $N_{agg}$ to be 0.96±0.01, which is essentially 1 within the experimental errors. This observation indicates the validity of our method in extracting the conformational properties of SDS micelles within this phase region.

Moreover, one can define the micellar boundary $R_b$ from $N_{agg}$ and $P(r)$ given in Figure 2(b) from the following equation

$$\int_0^{R_b} P(r) 4\pi r^2 dr = N_{agg}, (13)$$

$R_b$ is selected as the upper limit of the integration presenting on the LHS of Eqn. (13) to ensure that from the micellar center beyond this spatial threshold SDS molecules only exist as free unimers in water. Another commonly used parameter to measure the size of a micelle is radius of gyration $R_g$ which can be calculated from $P(r)$. From our analysis, $R_b$ and $R_g$ are estimated to be 28.2 Å and 17.1 Å, respectively. One can further calculate the intra-micellar volume fraction of SDS molecules $\phi_{SDS}$ and water $\phi_w$ by defining $\phi_w = v_w \cdot \frac{\int_0^l H(r) 4\pi r^2 dr}{\int_0^l 4\pi r^2 dr}$ and $\phi_{SDS} = 1 - \phi_w$. We use $R_b$ and $R_g$ as the upper limit of integrations $L$ to calculate $\phi_{SDS} = 0.84$ and $\phi_w = 0.16$, respectively. From the estimated $H(r)$ and $P(r)$ given in Figure 2, it can be concluded that additional invasive water molecules are mainly accommodated in the micellar periphery. As indicated by Figure 2, SDS molecules are highly localized around the micellar central region. Given this specific landscape of $P(r)$, the micellar spatial region defined by $R_g$, which is the second moment of $P(r)$, is mainly occupied by the dodecyl tail groups. Therefore, it is not surprising that the $\phi_{SDS}$ calculated based on $R_g$ is significantly higher than the corresponding $\phi_w$. One the other hand, from Eqn. (13), it is clearly seen that the water-rich micellar periphery is incorporated in the spatial region defined by $R_b$. The evolution of $R_b$ is therefore more correlated to the variation of intra-micellar hydration level.

In summary, we investigate the structure of SDS micelles in aqueous solutions using the contrast variation SANS technique. From the characteristic variation of SANS intensities caused by the change in the D/H ratio of water, we establish a methodology to experimentally determine the intra-micellar invasive water and SDS distributions, two crucial param-eters which allow us to quantitatively evaluate the conformational characteristics of SDS micelles. An important area for self-assembled systems in aqueous environments to be fully explored is the correlation between the degree of hydration and the self-organization of amphiphiles. Because the measured SANS spectra are sensitive to the isotopic ratio of water, it appears clear that our method, with adequate extensions to incorporate the effects such as non-spherical geometric shapes and non-negligible polydispersity, promises to further elucidate the micellization phenomenon from the perspective of conformational heterogeneity.

## ACKNOWLEDGMENT

This research at the SNS of the Oak Ridge National Laboratory was sponsored by the Scientific User Facilities Division, Office of Basic Energy Sciences, U.S. Department of Energy. K. C. acknowledges the support of the ORNL HSRE program.